\documentclass[pra,longbibliography,   amsmath, amssymb, 
  floatfix]{revtex4-2}
\usepackage{amsmath,graphics, graphicx, bm, natbib, color}
\usepackage[capitalise]{cleveref}
\usepackage{graphics, graphicx, bm, natbib, color, xcolor}
\usepackage{caption}
\usepackage{subcaption}
\usepackage{comment}
\usepackage{appendix}
\usepackage{mathtools}
\newcommand{\rem}[1]{}

\newcommand{\ud}{\mathrm{d}}
\newcommand{\diag}{\text{diag}}
\newcommand*{\vertbar}{\rule[-1ex]{0.5pt}{2.5ex}}

\newcommand{\writematrix}[1]{\mathbf{\underline{ #1}}}

\begin{document}

\author{Wijnand Broer}

\affiliation{School of Physical Sciences, University of Chinese Academy of Sciences, Beijing 100049, China}
\affiliation{School of Physics, Hefei University of Technology, Hefei, Anhui 230601,  China }
%

\author{Rudolf Podgornik}
\affiliation{School of Physical Sciences, University of Chinese Academy of Sciences, Beijing 100049, China}
\affiliation{Institute of Physics, Chinese Academy of Sciences, 100190 Beijing, China }
\affiliation{Kavli Institute for Theoretical Sciences, University of Chinese Academy of Sciences, Beijing 100049, China}
\affiliation{Wenzhou Institute of the University of Chinese Academy of Sciences, Wenzhou, Zhejiang 325000, China}\email{wbroer@gmail.com}
\altaffiliation{Department of theoretical physics, J. Stefan Institute,  1000 Ljubljana, Slovenia and Department of Physics, Faculty of Mathematics and Physics, University of Ljubljana, 1000 Ljubljana, Slovenia}
 \email{podgornikrudolf@ucas.ac.cn}

\title{Interplay between finite thickness and chirality effects on the Casimir-Lifshitz torque with nematic cholesteric liquid crystals.  }

\date{\today}

\begin{abstract}
We theoretically investigate the combined effects of the chirality and the finite total thickness of nematic cholesteric liquid crystals on the Casimir-Lifshitz torque. We find that, the larger the thickness, the more sinusoidal the angular dependence of the torque becomes. We use a Fourier decomposition to quantify this result. The general direction of the torque depends on whether the configuration of two cholesterics is heterochiral or homochiral.
\end{abstract}
\maketitle
\section{Introduction}

The Casimir effect \cite{Casimir48,Lifshitz55,Lifshitz61} is a macroscopic dispersion force that originates from the quantum mechanical and thermal fluctuations of the electromagnetic (EM) field. The term 'dispersion forces' refers to the fact that their properties are governed by the electric and magnetic susceptibilities of the materials involved \cite{vdWBook,CasimirBook, BuhmannBookI, Woods2016Review}. The name 'van der Waals-London force' is associated with the microscopic or non-retarded version of the Casimir force \cite{vdWBook}.

Casimir forces are investigated for both practical and fundamental reasons. The fundamental motivation pertains to the search for hypothetical new forces and deviations from Newtonian gravity in short range gravitation measurements \cite{Sedmik2021} More practically, Casimir interactions affect the actuation dynamics of nano- and micro-mechanical systems, such as switches, cantilevers, and actuators at a sub-micrometer length scale \cite{Serry1995,Chan2001, Broer2015, Mehdi2015, Tajik2018PRE} as MEMs devices are commonly used in Casimir force experiments \cite{Munday2021Review}. Recent developments in the field of quantum sensing \cite{Degen2017} may open up the possibility for new techniques to measure Casimir forces and torques. Moreover,  the Casimir interactions underpin the stability of colloidal and biophysical macromolecular systems \cite{FrenchReview2010}. These forces represent one of the pillars of the fundamental Deryaguin-Landau-Verwey-Overbeek theory of colloid stability \cite{Israelachvili}. 

Casimir force calculations require knowledge of the electric and magnetic susceptibilities as a function of frequency in a broad range\cite{Lifshitz55,*Lifshitz61,Woods2016Review}. Furthermore, the Casimir force depends on the shapes of the interacting bodies in a non-trivial way, because its evaluation requires solving the Maxwell equations for the given geometry\cite{BuhmannBookI,Johnson2011}. 

Both the material properties and the shapes can exhibit anisotropy \cite{Hopkins2015}, providing the bodies with a well defined orientation. If this anisotropy lies in the plane reflection, the Casimir interaction can manifest as a torque that attempts to align the orientations of the bodies, the so callled Casimir torque \cite{Parsegiantorque}. In what follows we will limit ourselves to planar geometries with dielectric (i.e. material induced) anisotropy. A full analytical description of the Casimir torque between birefringent half spaces was based on the direct solution of the Maxwell equations and the pertaining dispersion equation \cite{Barash1978}. Alternatively, the Maxwell equations can be solved by interpreting them as eigenvalue problem. This method is known as the transfer matrix method. It is especially suited for the description of electromagnetic wave propagation through anisotropic media \cite{Berreman1972,Yeh1979}. Some recent examples of the application of the transfer matrix method in the context of the Casimir torque can be found in Refs. \cite{Zeng2016,Thiyam2018,Emelianova2020,Farias2020,Zeng2020,ChenLiang2020}. It has been shown \cite{Broer2019} that the result of the transfer matrix method is consistent with the more direct method using the dispersion equation from Ref. \cite{Barash1978}. The first experimental observation of the Casimir torque \cite{Somers2018} used a liquid crystal in the nematic phase, that effectively behaves like a birefringent half space \cite{Somers2015}. 

Here we will focus on cholesteric liquid crystals, which is a different phase. Unlike nematic liquid crystals, cholesterics consist of many thin layers where each layer has a slightly different orientation than the adjacent one. Macroscopically, cholesterics exhibit a chiral helical structure \cite{Ryabchun2018}. Chirality refers to a lack of the mirror symmetry associated with the handedness of the helix. Handedness has the symmetry properties of a pseudo-scalar and can only have two values, differentiating between right and left handed helices. It behaves as a scalar in all symmetry transformations, except under a parity inversion where it changes sign.  Previously \cite{Broer2021} we introduced a model to describe the effect of this helical shape on the Casimir torque. Here we expand this model to include finite thickness effects as well.  Both effects will be taken into account in correlation. This requires a non-trivial regularization of the transfer matrix to prevent exponential divergences at the imaginary Matsubara frequencies.

The paper is organized as follows. After the introduction we will discuss the used methodologies: the Lifshitz theory and what we call the spiral staircase model for cholesterics. Next we will present the numerical results obtained from these methods. Finally we will summarize and present an outlook.

\section{Formalisms and methodologies}

\subsection{Lifshitz formula}

 {\sl Grosso modo} one can distinguish two approaches to theoretically describe Casimir-Lifshitz forces. Firstly, the approach based on the Green function tensors is closest to the original papers by Lifshitz et al \cite{Lifshitz55,*Lifshitz61}. (See e.g. Refs. \cite{Schwinger78, Lambrecht2006, Scheel2008}). Alternatively,  the  derivation based on the summation over the allowed EM modes  within a given geometry \cite{vanKampen68, Schram1973, Barash1984} is  more akin to the original paper by Casimir \cite{Casimir48}. It can be shown that both approaches lead to the same result, \cite{Intravaia2012} namely the Lifshitz formula for dispersion interactions.

According to the Lifshitz formula, the free energy per unit area is given by: 

\begin{equation}
 \frac{E_{Cas}}{A}=\frac{k_b T }{4\pi^2}\sum_{n=0}^{\infty}(1-\tfrac{1}{2}\delta_{0n})\int\limits_0^\infty\int\limits_0^{2\pi}\ln[D(k_\rho,\eta,\{\theta_j\}, \varphi, i\zeta_n)]k_\rho\ud k_\rho\ud\eta
\label{eq:Lifshitz}
 \end{equation}
where \cite{Lambrecht2006}
\begin{equation}
 D=\det(\underline{\pmb{I}}-\underline{\pmb{r}}_1(\theta_1, i\zeta_n)\cdot\underline{\pmb{r}}_2(\theta_1+\varphi, i\zeta_n)e^{-2k_3a}).
\end{equation}
Here $k_\rho$ and $\eta$ denotes the radial and azimuthal components of the wave vector.
And the $2\times2$  reflection matrices 

\begin{equation}
 \underline{\pmb{r}}_q=\begin{pmatrix}
                        r_{q,ss}&r_{q,sp}\\
                        r_{q,ps}&r_{q,pp}
                       \end{pmatrix} \qquad q\in \{1,2\},
\end{equation}
are obtained by solving the Maxwell equations, as we will outline in what follows. Even though the matrices $\underline{\pmb{r}}_1$ and $\underline{\pmb{r}}_2$ in \cref{eq:Lifshitz}  generally do not commute with each other, it follows from the Sylvester determinant identity\cite{PozrikidisBook} that 

\[\det(\underline{\pmb{I}}-\underline{\pmb{r}}_1(\theta_1, i\zeta_n)\cdot\underline{\pmb{r}}_2(\theta_1+\varphi, i\zeta_n)e^{-2k_3a})=\det(\underline{\pmb{I}}-\underline{\pmb{r}}_2(\theta_1+\varphi, i\zeta_n)\cdot\underline{\pmb{r}}_1(\theta_1, i\zeta_n)e^{-2k_3a}),\]
so that the multiplication order will not affect that Casimir energy. This is to be expected physically because the arbitrary choice of which surface is located to the 'left' and which to the 'right' should not affect the result. 

All quantities are evaluated at the imaginary Matsubara frequencies $ \zeta_n = \frac{2\pi n k_b T}{\hbar},$ so that each contribution to the Casimir energy decreases monotonically with $n$. The Casimir torque is then given by 

\begin{equation}\label{eq:torque}
 \tau(a,\varphi)=-\frac{\partial E_{Cas}}{\partial \varphi} 
\end{equation}
where $\varphi$ denotes the  angle between the optic axes of the layers of each stack closest to the gap.

To determine the Fresnel reflection matrices, we will use the transfer matrix method, outlined in the appendix.

\subsection{Spiral staircase model for cholesterics}

The transfer matrix method is applicable to all anisotropic planar multilayer structures. For examples of its application to Casimir interactions, see Refs. \cite{Veble2009,Zeng2016,Broer2019} However, the spiral staircase model has been specifically designed for cholesterics. Here we will provide its main analytical results. For more details see Ref. \cite{Broer2021}.

 The main idea behind the spiral staircase model is to combine two approximations: 1) the application of the Baker-Campbell-Haussdorff (BCH) \cite{HighamBook2008}  formula, and 2) Approximating the discrete layers as a continuously varying function of $z$, by means of finite difference formulas \cite{Fornberg1988}.

 The BCH  formula for two adjacent layers $\writematrix{Q}_1$ and $\writematrix{Q}_2$ each with thickness $d$  is
 
 \begin{equation}
        \exp({-i\underline{\pmb{Q}}_1d})\exp({-i\underline{\pmb{Q}}_2d})=\exp\left({-i(\underline{\pmb{Q}}_1+\underline{\pmb{Q}}_2)d}-\frac{d^2}{2}[\underline{\pmb{Q}}_1,\underline{\pmb{Q}}_2]+O(d^3)\right),
        \label{eq:BCH2layers}
 \end{equation}
 which is valid if the thickness of each layer is small on the scale of the matrix norm of $\writematrix{Q}_{1,2}$:
    
    \begin{equation}\label{eq:BCHCondition}
        ||\writematrix{Q}_{1,2}||d\ll1.
    \end{equation}
Physically, \cref{eq:BCHCondition} is justified by the fact that the layers of the cholesteric are of the atomic or molecular length scale, whereas the Lifshitz theory operates on a continuous medium length scale. Hence $d$ can be considered infinitesimally small. 

\begin{figure}[!htbp]
\centering
\includegraphics[width=\linewidth]{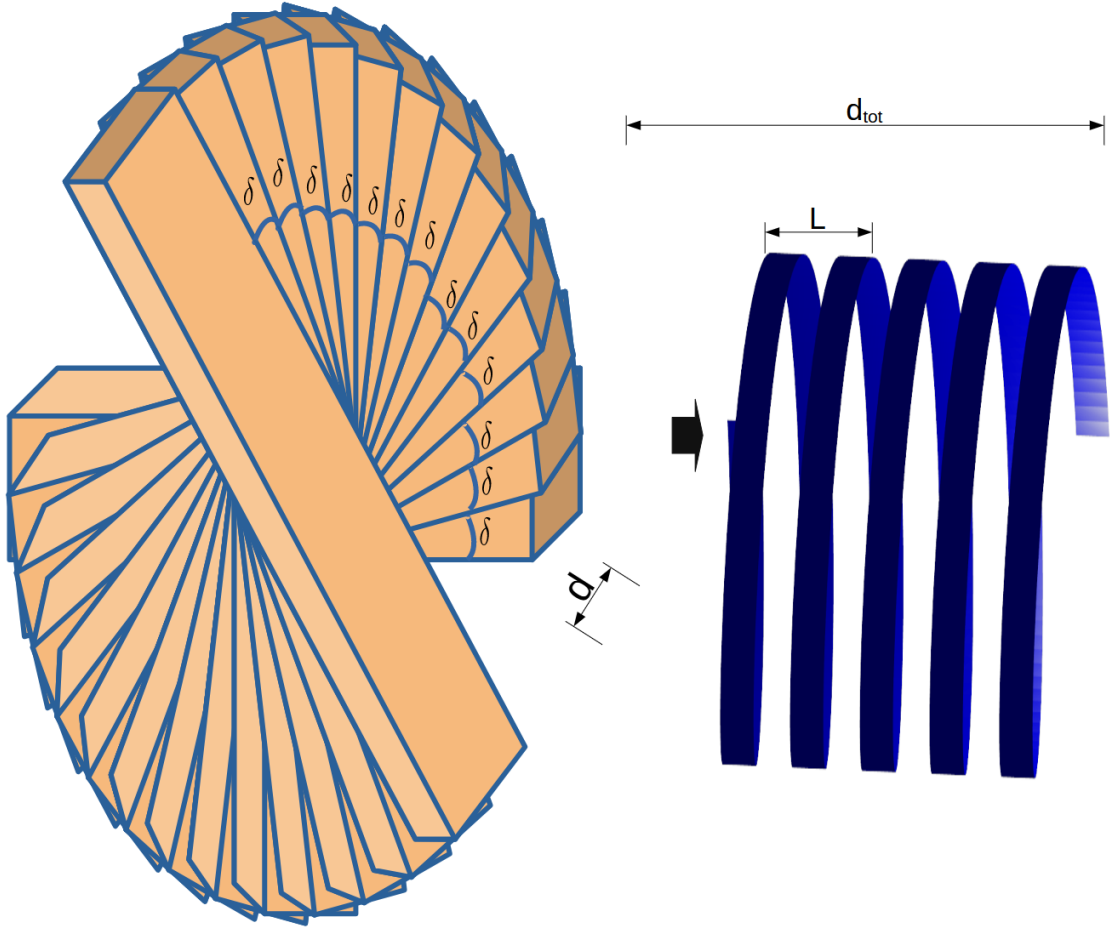}
 \caption{Illustration of the spiral staircase model. Infinitesimally thin layers each rotate by a small amount $\delta$ with respect to the adjacent layer. This planar multilayer geometry can be approximated as a continuous helix, which models a cholesteric. The error of this approximation is of the order $\delta^2$. }
 \label{fig:SpiralStaircase}
\end{figure}

We assume that each layer has the same thickness $d$ (in the order  of nm), and the cholesteric has a helical shape on a macroscopic length scale. The layers are identical in every respect, except that each layer differs in orientation by an angle $\delta\ll1$ from the adjacent layer. (See \cref{fig:SpiralStaircase}). So the \emph{pitch length} $L$ of the helix becomes

\begin{equation}
    L=\frac{\pi}{\delta}d.
\end{equation}
Since $L$, $\delta$, and $d$ are assumed to be constant one can choose  to define the BCH expansion in $\delta$ instead of $d$
 \begin{equation}
        \exp({-i\underline{\pmb{Q}}(\theta_1)_1d})\exp({-i\underline{\pmb{Q}}(\theta_1+\delta)_2d})=\exp\left({-i(\underline{\pmb{Q}}_1+\underline{\pmb{Q}}_2)\frac{L\delta}{\pi}}+O(\delta^2)\right),
        \label{eq:BCH2delta}
 \end{equation}
where we have lowered the order of the BCH expansion. This leading order contribution represents the effect of the curvature of the helix. The next order term signifies the effect of the end of the pitch. It has turned out that the latter is negligible at separations larger than 100 nm \cite{Broer2021}. The continuity approximation can be expanded up to the same order:

\begin{equation}
    \underline{\pmb{Q}}_{j+1}((j+1)\delta)=\underline{\pmb{Q}}_j+\delta\frac{\partial \underline{\pmb{Q}}_j }{\partial \theta}\left(j\delta\right)+O(\delta^2)\quad j=1,2,3\text{...}
\end{equation}
Note that the order can be increased by adding more layers within the expansion, but now there is a self consistent approximation in $\delta$.

Therefore we will limit ourselves to the leading order BCH approximation, where the commutator of the transfer matrices of two adjacent layers is neglected. It can be shown that this leads to an additive approximation where the extraordinary eigenvalues are averaged \cite{Broer2021}.So we start with the integral

\begin{equation}\label{eq:qint}
    q_{\text{int}}=\frac{L}{d_{tot}\pi}\int\limits_0^{d_{tot}\pi/L}q_e(\theta)\ud\theta,
\end{equation}
where $q_e$ is the extraordinary eigenvalue given by the function

\begin{equation}\label{eq:qe_function}
q_{e}(\theta)= \sqrt{\varepsilon_{x}\zeta^2/c^2+(\varepsilon_{x}/\varepsilon_{y})k_\rho^2\cos^2(\theta)+k_\rho^2\sin^2(\theta)}    
\end{equation}
 Since the total thickness $d_{tot}$ is finite here we can no longer assume that the number of 'pitches' (periods) is an integer. Therefore the integral in \cref{eq:qint} can no longer be calculated analytically. The averaged orientation is obtained through the inverse of \cref{eq:qe_function}:

\begin{equation}\label{eq:thavg}     \left<\theta\right>=\pm\arccos\left(\pm\frac{\sqrt{k_\rho^2+\varepsilon_{1x}\zeta^2/c^2-q_{\text{int}}^2}}{k_\rho\sqrt{1-\tfrac{\varepsilon_{1x}}{\varepsilon_{1y}}}}\right). 
\end{equation}
Depending on the sign of $\left<\theta\right>$, the crystal has left handed ('$-$') or right handed ('$+$')  chirality. The signs within the argument of the arc-cosine of \cref{eq:thavg} corresponds to the propagation direction of the wave: the plus sign represents a forward propagating wave, and a minus sign represents a backward propagating wave. Using \cref{eq:qint,eq:thavg} we can determine the transfer matrix for a single slab with finite thickness:

\begin{equation}    \writematrix{T}_1=\left<\writematrix{S}_1\right>\cdot\left<\writematrix{P}\right>\cdot\left<\writematrix{S}_1^{-1}\right>+O(\delta^2)
\end{equation}
where 

\begin{equation}    \left<\writematrix{S}_1\right>=\writematrix{S}_1(\left<\theta\right>),\quad \left<q_e\right>\equiv q_e(\left<\theta\right>)
\end{equation}
and

\begin{equation}\label{eq:Pavg}
    \left<\writematrix{P}\right>=\diag\left(\exp(\left<q_{e}\right>d_{tot}),\exp(-\left<q_{e}\right>d_{tot}),\exp(q_{o}d_{tot}),\exp(-q_{o}d_{tot})\right),
\end{equation}
where the arguments of the exponential functions are real at imaginary frequencies.

The total transfer matrix is
\begin{equation}\label{eq:MIdeal}
    \writematrix{M} = \writematrix{S}_0^{-1}\left<\writematrix{S}_1\right>\cdot\left<\writematrix{P}\right>\cdot\left<\writematrix{S}_1^{-1}\right>\writematrix{S}_0+O(\delta^2)
\end{equation}

Once the total transfer matrix is known, the Fresnel matrix elements are given by \cite{Yeh1979,Passler2017}:

\begin{subequations}\label{eq:Fresnel}
\begin{gather}
r_{ss}=\frac{M_{21} M_{33}-M_{23} M_{31}}{M_{11} M_{33}-M_{13} M_{31}}
\label{eq:rss}\\
r_{sp}=\frac{M_{33} M_{41}-M_{31} M_{43}}{M_{11} M_{33}-M_{13} M_{31}}
\label{eq:rsp}\\
r_{ps}=\frac{M_{11} M_{23}-M_{13} M_{21}}{M_{11} M_{33}-M_{13} M_{31}}
\label{eq:rps}\\
r_{pp}=\frac{M_{11} M_{43}-M_{13}M_{41}}{M_{11} M_{33}-M_{13} M_{31}}.
\label{eq:rpp}
\end{gather}
\end{subequations}
While \cref{eq:MIdeal} is straightforward to understand from a mathematical point of view, depending on the physical system, it can be difficult to implement in practice. We will address this in more detail in the next section.

\section{Results and discussion}\label{sec:Disc}


\begin{figure}[!htbp]
\begin{subfigure}[t]{0.45\textwidth}
\centering
\includegraphics[width=\linewidth]{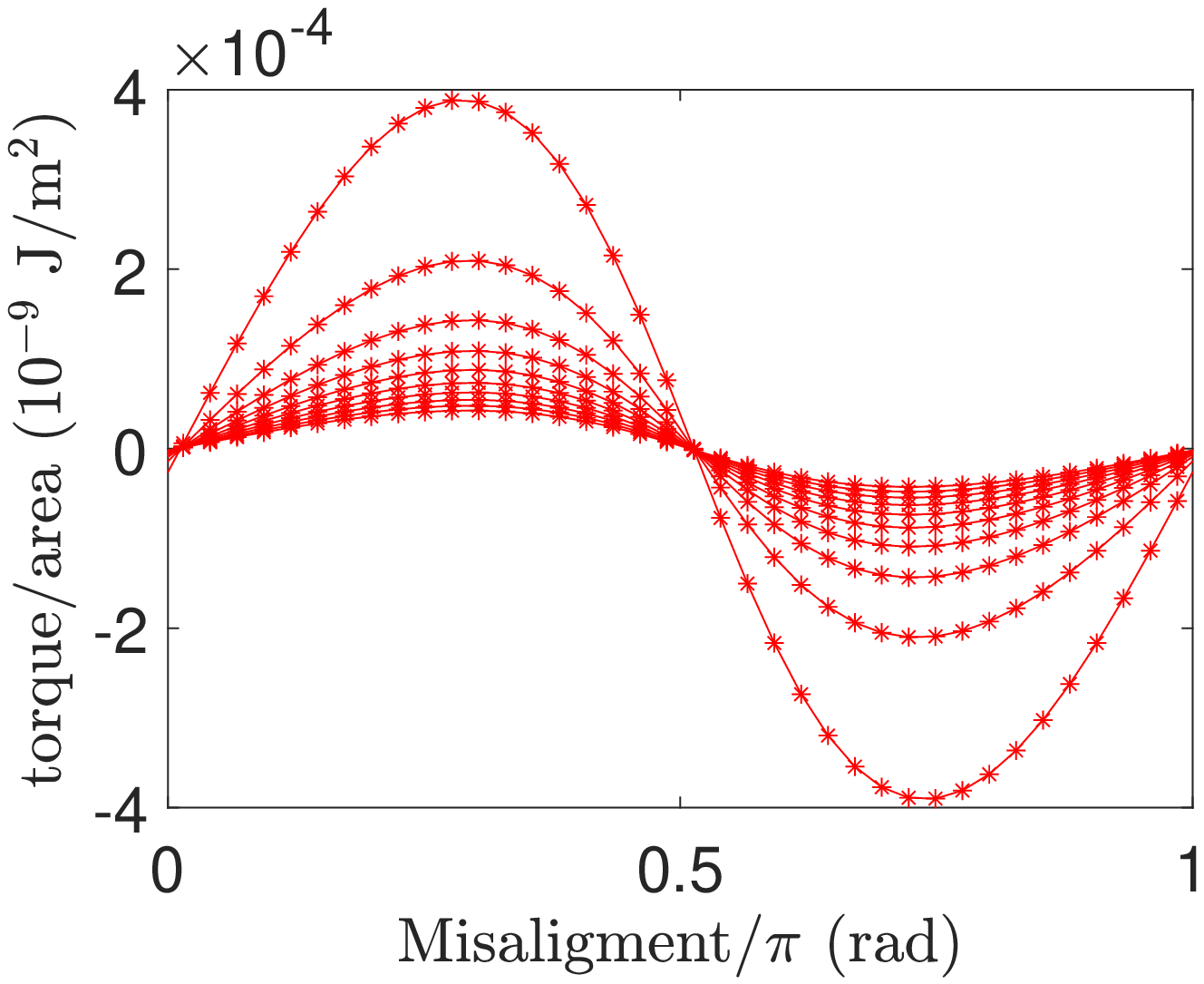}
 \caption{ }
 \label{fig:hom515}
\end{subfigure}
\hfill
\begin{subfigure}[t]{0.45\textwidth}
\centering
\includegraphics[width=\linewidth]{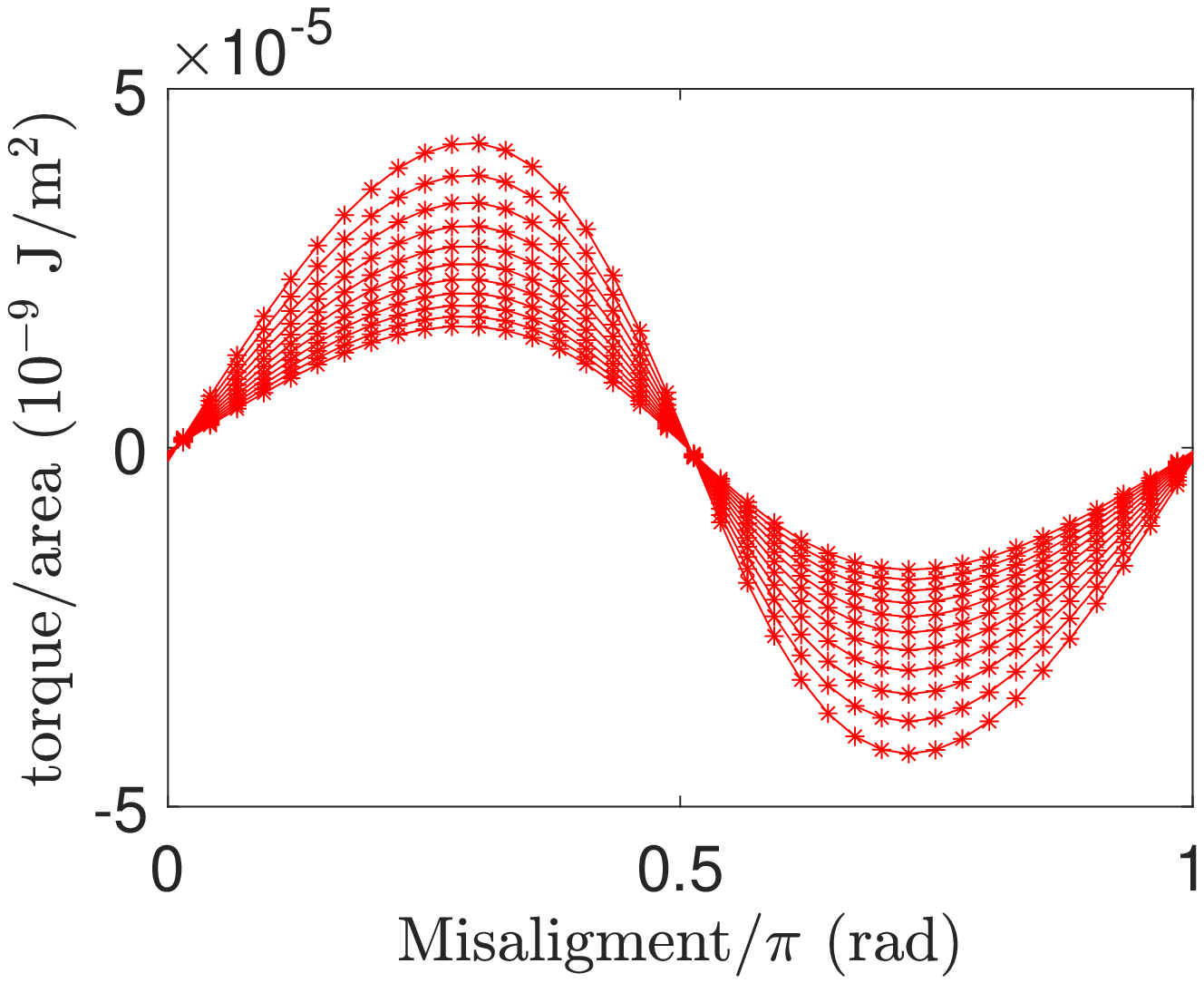}
 \caption{ }
 \label{fig:hom5510}
\end{subfigure}
\hfill
\begin{subfigure}[t]{0.45\textwidth}
\centering
\includegraphics[width=\linewidth]{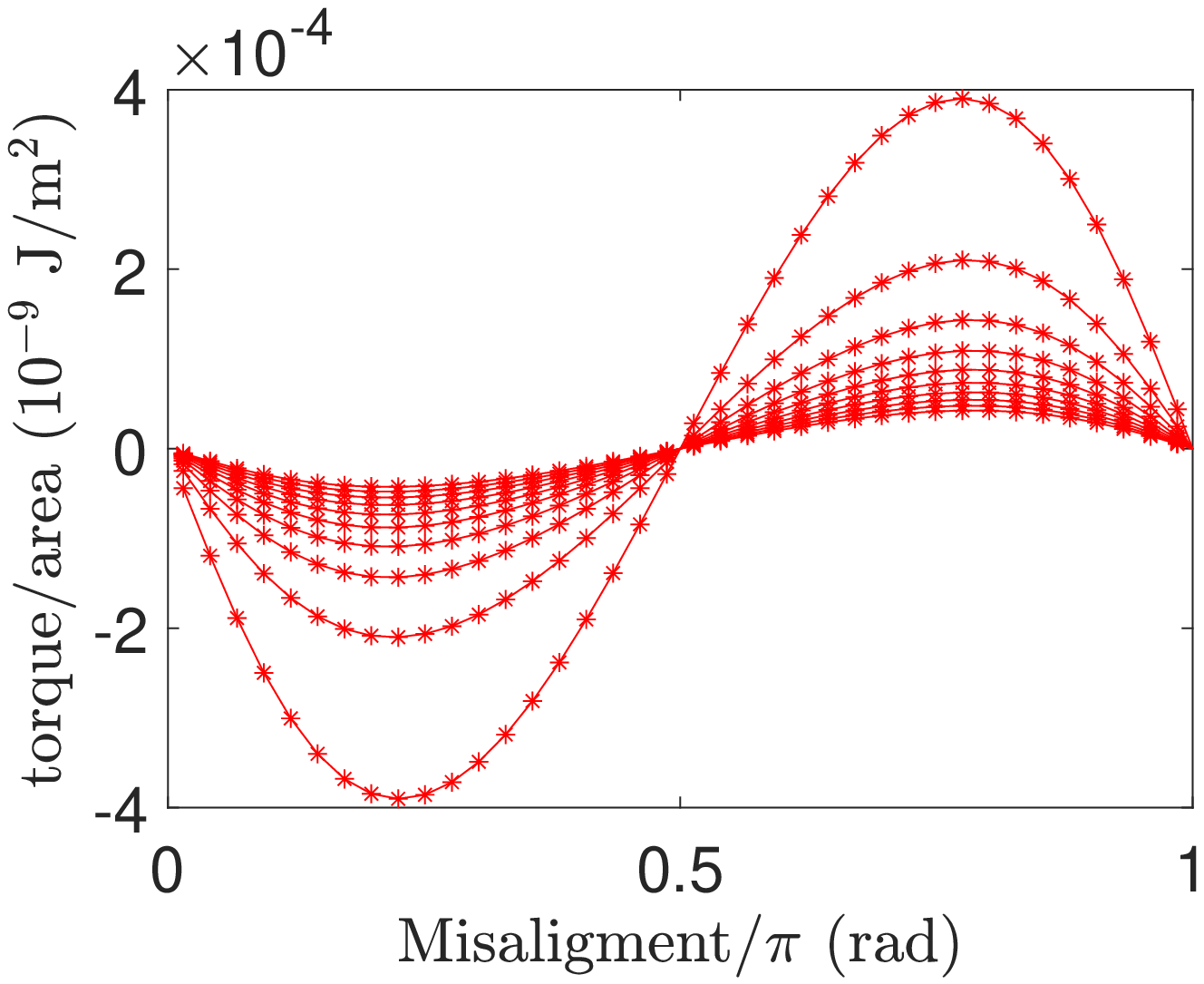}
 \caption{ }
 \label{fig:het515}
\end{subfigure}
\hfill
\begin{subfigure}[t]{0.45\textwidth}
\centering
\includegraphics[width=\linewidth]{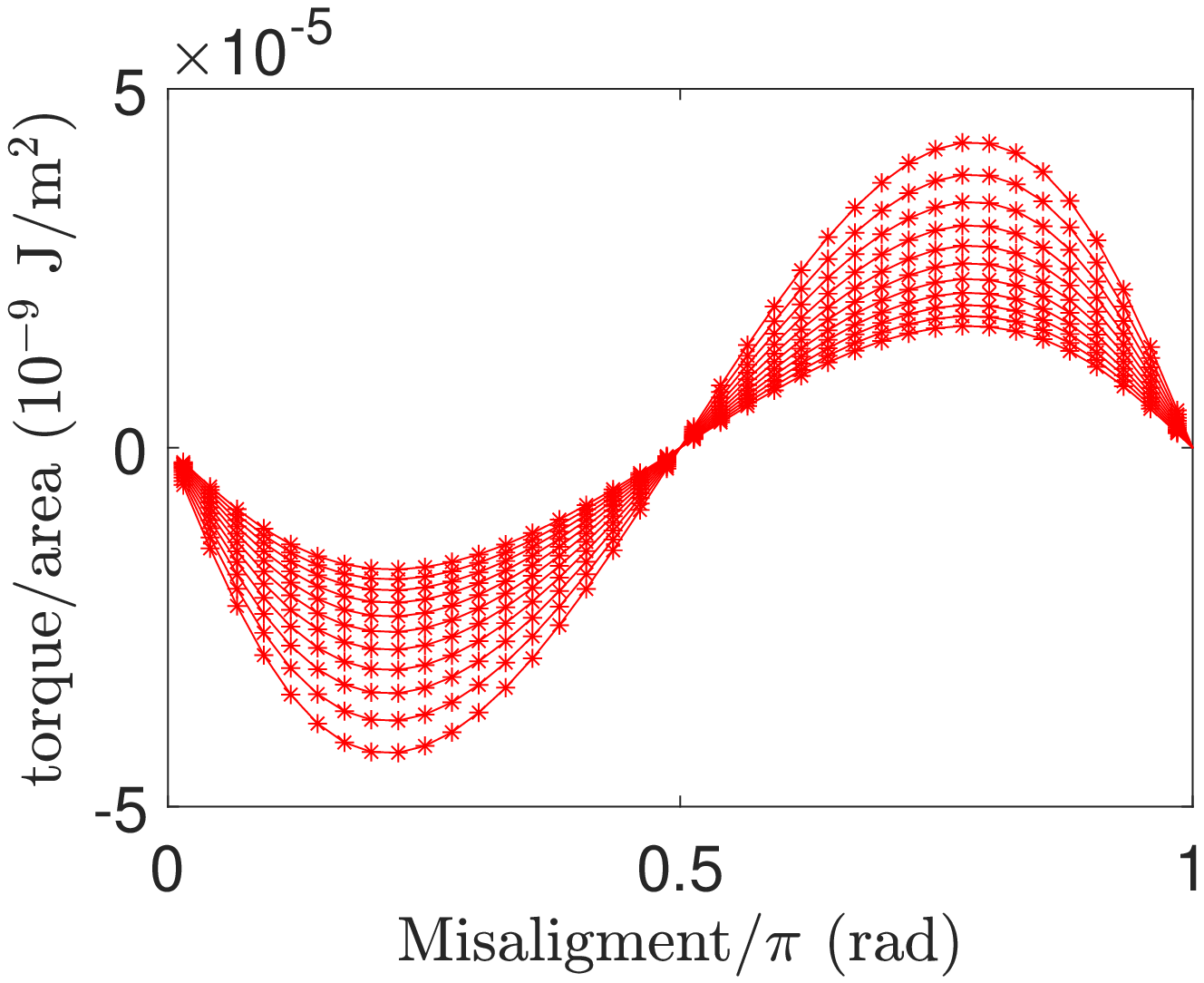}
 \caption{ }
 \label{fig:het5510}
\end{subfigure}
 \caption{The Casimir torque for two cholesteric crystals with thicknesses of 5 microns. Panel (a):  Homochiral configuration, separations between 1 and 5 microns, (b): Homochiral configuration, separations between 5 and 10 microns.  (c): Heterochiral configuration, separations between 1 and 5 microns,  (d):  Heterochiral configuration, separations between 5 and 10 microns.}\label{fig:5MicronTorque}
\end{figure}

\begin{figure}[!htbp]
\begin{subfigure}[t]{0.45\textwidth}
\centering
\includegraphics[width=\linewidth]{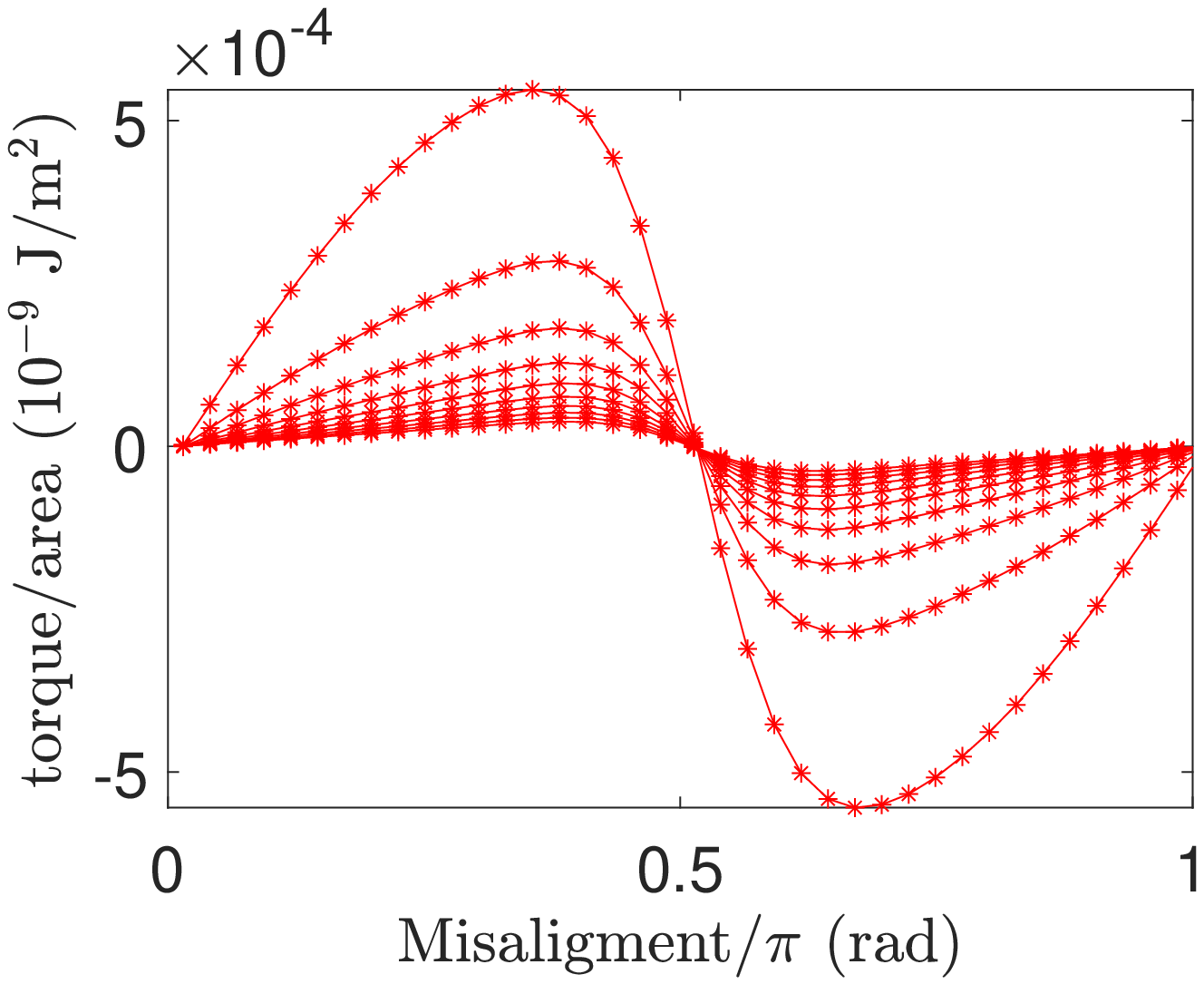}
 \caption{ }
 \label{fig:Hom11}
\end{subfigure}
\begin{subfigure}[t]{0.45\textwidth}
\centering
\includegraphics[width=\linewidth]{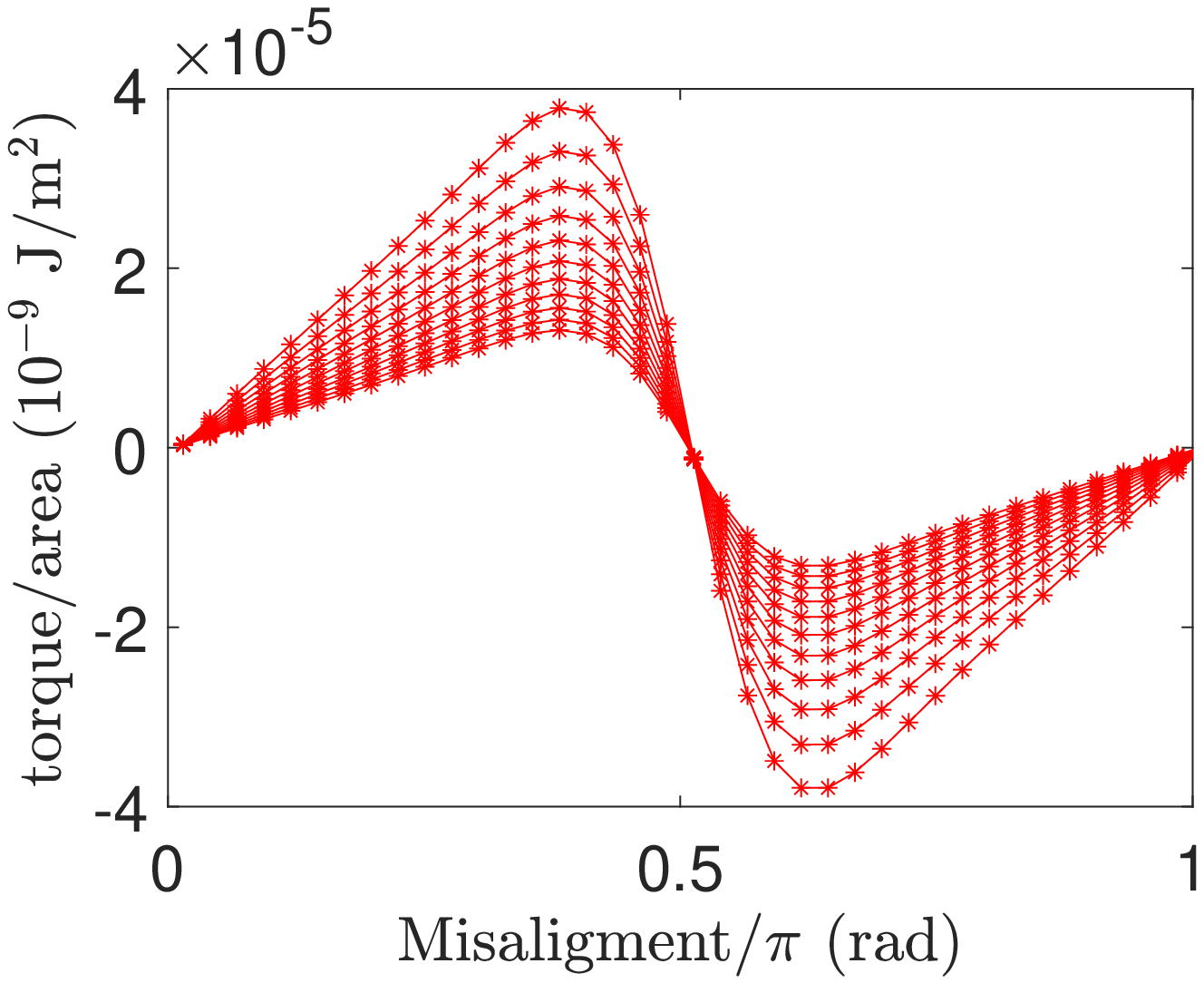}
 \caption{ }
 \label{fig:6}
\end{subfigure}
\begin{subfigure}[t]{0.45\textwidth}
\centering
\includegraphics[width=\linewidth]{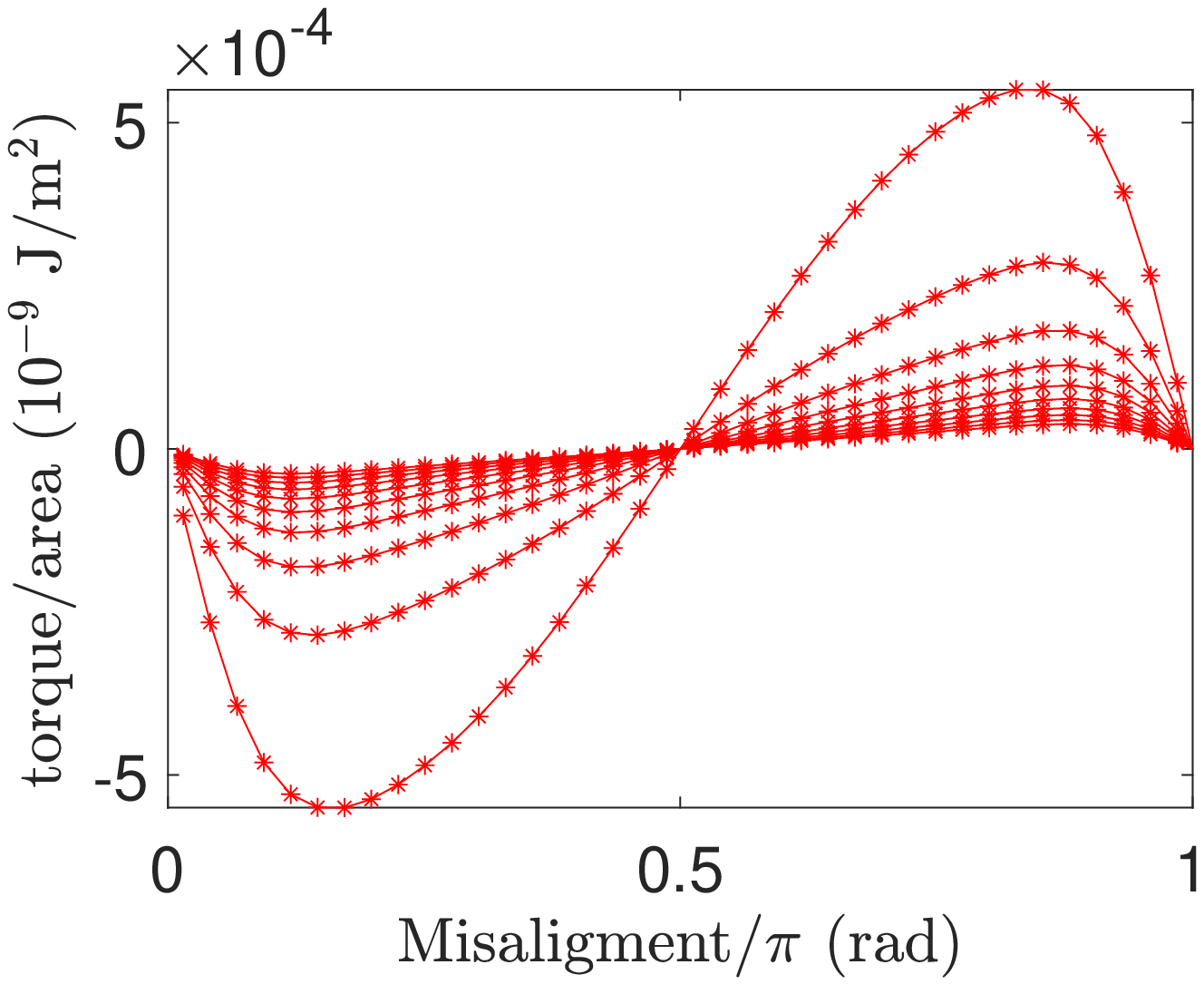}
 \caption{ }
 \label{fig:Het11}
\end{subfigure}
\begin{subfigure}[t]{0.45\textwidth}
\centering
\includegraphics[width=\linewidth]{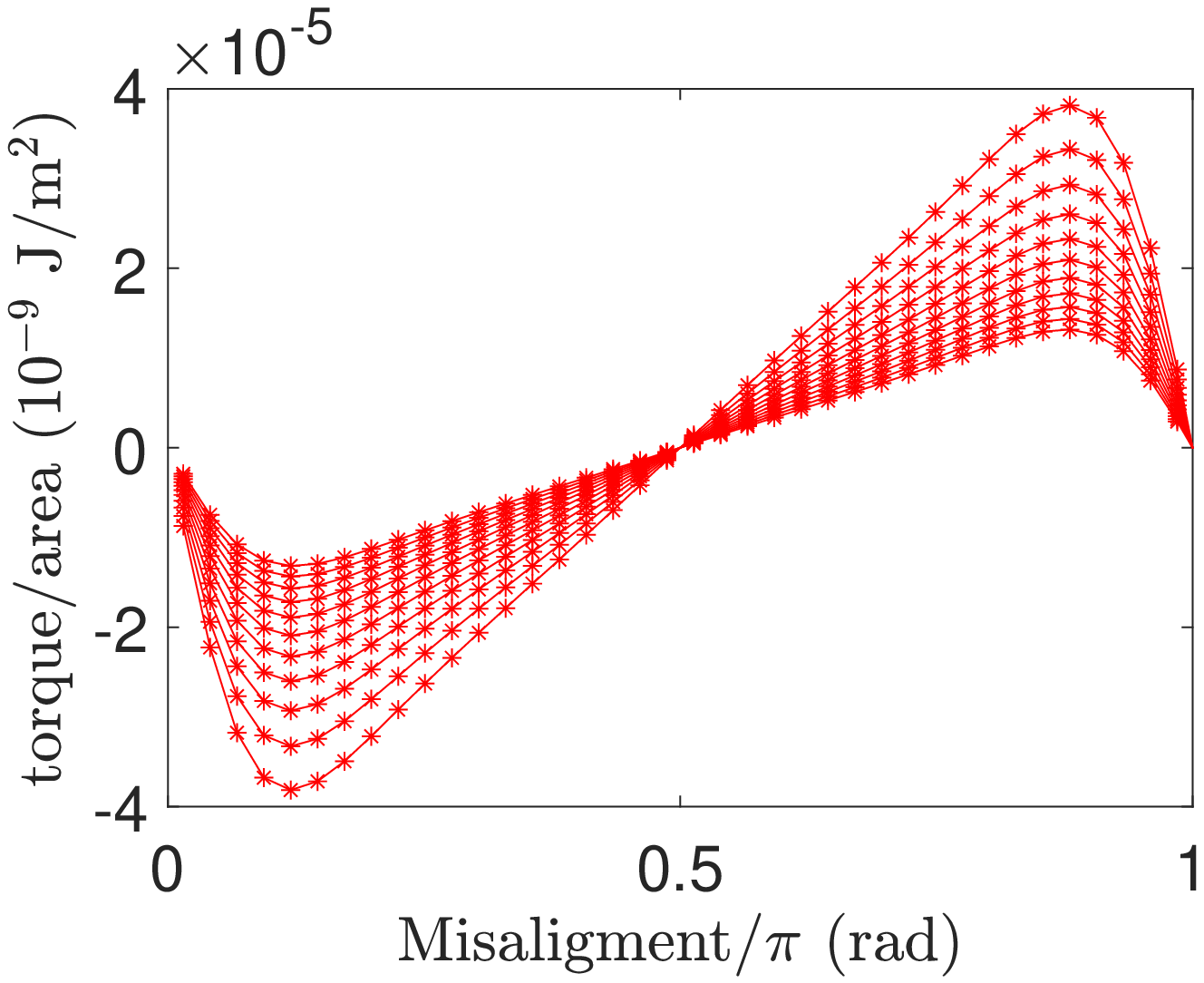}
 \caption{ }
 \label{fig:Het12}
\end{subfigure}
 \caption{The Casimir torque for two cholesteric crystals with thicknesses of 1 micron. Panel (a):  Homochiral configuration, separations between 1 and 5 microns, (b): Homochiral configuration, separations between 5 and 10 microns.  (c): Heterochiral configuration, separations between 1 and 5 microns,  (d):  Heterochiral configuration, separations between 5 and 10 microns. }\label{fig:1MicronTorque}
\end{figure}

Here we will numerically implement the Casimir torque given by \cref{eq:torque} with the reflection matrices given by \cref{eq:Fresnel} and the transfer matrix given by \cref{eq:MIdeal} One problem is the transition from anisotropic to isotropic materials, which can be discontinuous because in the isotropic case, the modes are degenerate\cite{Xu2000,Passler2017}. (The extraordinary eigenvalues become ordinary). However, since the Casimir torque is a pure anisotropy effect, this is not a concern for this kind of calculations.

However, numerically, \cref{eq:MIdeal} can lead to exponential divergences at large frequencies since the eigenvalues asymptotically increase linearly with frequency. (See \cref{eq:Pavg}) More precisely, for the numerical implementation we have to ensure that

\begin{equation}\label{eq:MReal}
     \writematrix{M} =\begin{cases}                                                    \writematrix{S}_0^{-1}\left<\writematrix{S}_1\right>\cdot\left<\writematrix{P}\right>
     \cdot\left<\writematrix{S}_1^{-1}\right>\writematrix{S}_0\quad&\text{if } \max(\exp(\left<q_{e}\right>d_{tot}),\exp( q_o d_{tot}))<\infty\\
                    \writematrix{S}_0^{-1}\left<\writematrix{S}_1\right>\quad&\text{otherwise,}         
                        \end{cases}
\end{equation}
where higher order terms in $\delta$ are neglected. This is because the scale of the thickness is  determined by the eigenvalues. Specifically, the limit of $\max(\left<q_{e}\right>d_{tot}), q_o d_{tot})\gg1$ corresponds to an \emph{effective} infinite thickness, where the thickness is much larger than the inverse of the smallest eigenvalue. Note that such divergences will also occur without chirality. Even though calculations for finite thickness anisotropic plates exist (e.g. Ref. \cite{Zeng2020}), we have not seen the divergences explicitly addressed anywhere. 

We will use the same dielectric tensor as in Ref. \cite{Broer2021} .  Let the cholesteric liquid crystal consist of 96 \% nematic 5CB (4- cyano-4'-pentyl-biphenyl) doped with 4 \% chiral dopant S811 by mass. \cite{Kocakuelah2021} The value of the static Debye term for this mixture was taken from a recent experiment \cite{Kocakuelah2021}. The dielectric function of 5CB  was established in Ref. \cite{Kornilovitch2012} based on data from Ref. \cite{Wu1993}:

\begin{equation}\label{eq:eps_tot}
\varepsilon_{x,y}(i\zeta)\approx\varepsilon_{D,x,y}(0)+\varepsilon_{5CB,x,y}.(i\zeta)
\end{equation}

We repeat that at this range of separations, in the order of several microns, only the leading order term of the BCH expansion needs to be included. Consequently the exact value of the pitch length doesn't affect the Casimir torque. It is assumed that at least one pitch fits inside the crystal, i.e.

\begin{equation}\label{eq:1rotation}
    L<d_{tot}
\end{equation}
in other words, the helix completes at least one rotation. Physically, the lowest order term in the BCH expansion represents the contribution due to the curvature of the helix, the higher order term contributes only at separations less than 100 nm, where the Casimir torque can 'feel' the end of the pitch\cite{Broer2021}. Therefore the exact value of $L$ doesn't matter as long as the condition \cref{eq:1rotation} is met.

To get a sense of the how the finite thickness effects manifest, we have performed calculations for crystals with thicknesses of 1 and 5 microns. For simplicity we take both interacting crystals to be of the same thickness, so that there is only one thickness to consider in each case, $d_{tot}$.  

The results for the Casimir torque for a thickness of five microns are displayed in \cref{fig:5MicronTorque} at separation distances between 1 and 10 microns. Since the separation distances are comparable to the thickness, some finite thickness effects may be expected. However, the shape of the curves remains largely sinusoidal. (We will later define this more quantitatively). It is worth noting that the direction of the torque depends on whether the crystals have the same chirality (referred to as the homochiral case) or a different chirality (referred to as the heterochiral case).

\cref{fig:1MicronTorque} shows the Casimir torque for the 1 micron thick crystals, where we expect the finite thickness effects to be more pronounced. We see a similar trend as in the previous case, that is, the general direction depends on whether the configuration is heterochiral or homochiral. However, in this case the torques do not differ simply by a minus sign from each other, because the curves are no longer symmetric around $\varphi=\pi/4$ and $\varphi=3\pi/4$. Indeed the curves significantly deviate from a sinusoidal pattern. This deviation becomes even more pronounced, at larger distances which suggests that the finite thickness effects are associated with it. Furthermore, the amplitude of the torque is larger than in the previous case at the short distances of 1 or 2 microns, but it decreases faster as a function of distance.  

\begin{figure}[!htbp]
\begin{subfigure}[t]{0.45\textwidth}
\centering
\includegraphics[width=\linewidth]{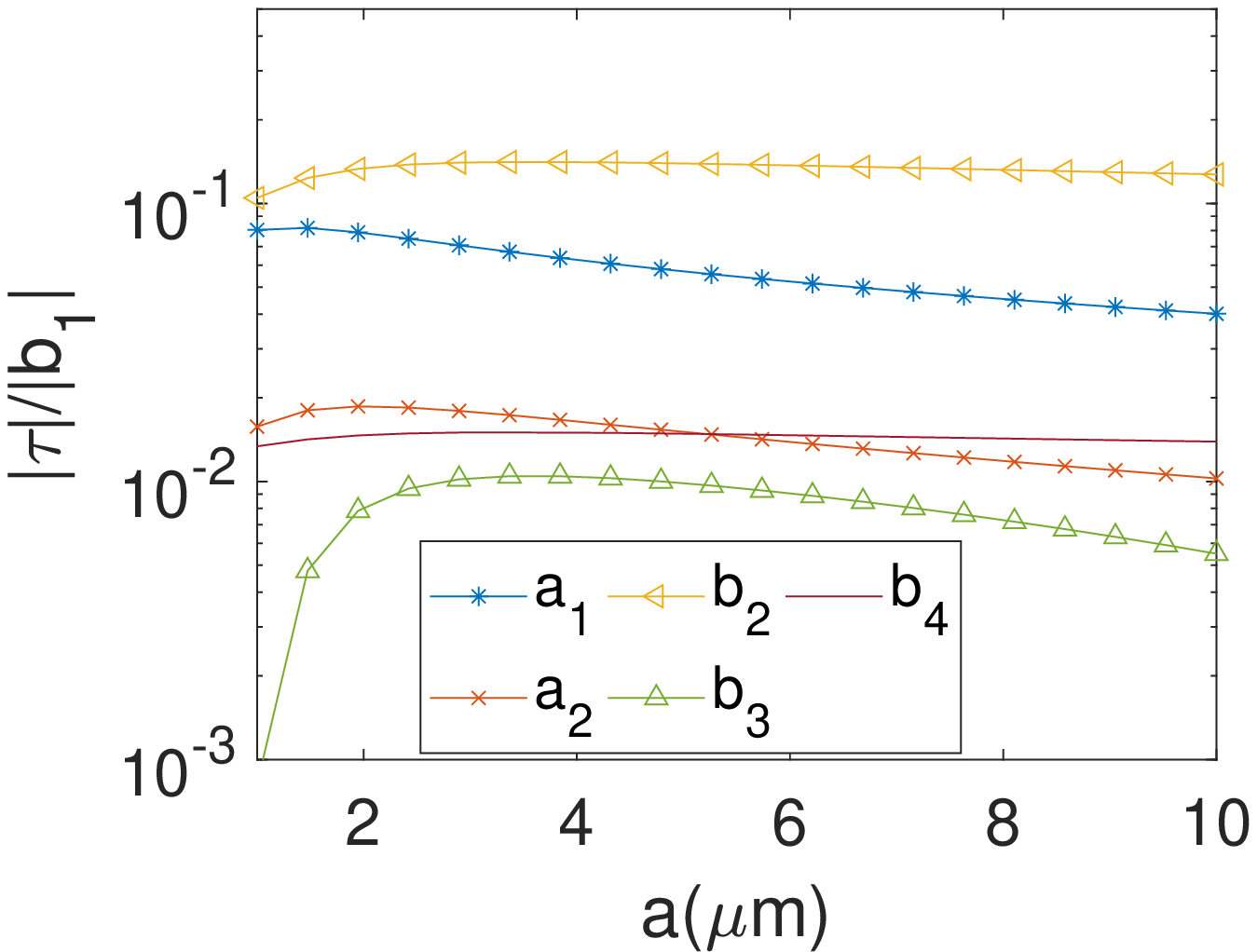}
 \caption{ }
 \label{ig:Fourier51}
\end{subfigure}
\begin{subfigure}[t]{0.45\textwidth}
\centering
\includegraphics[width=\linewidth]{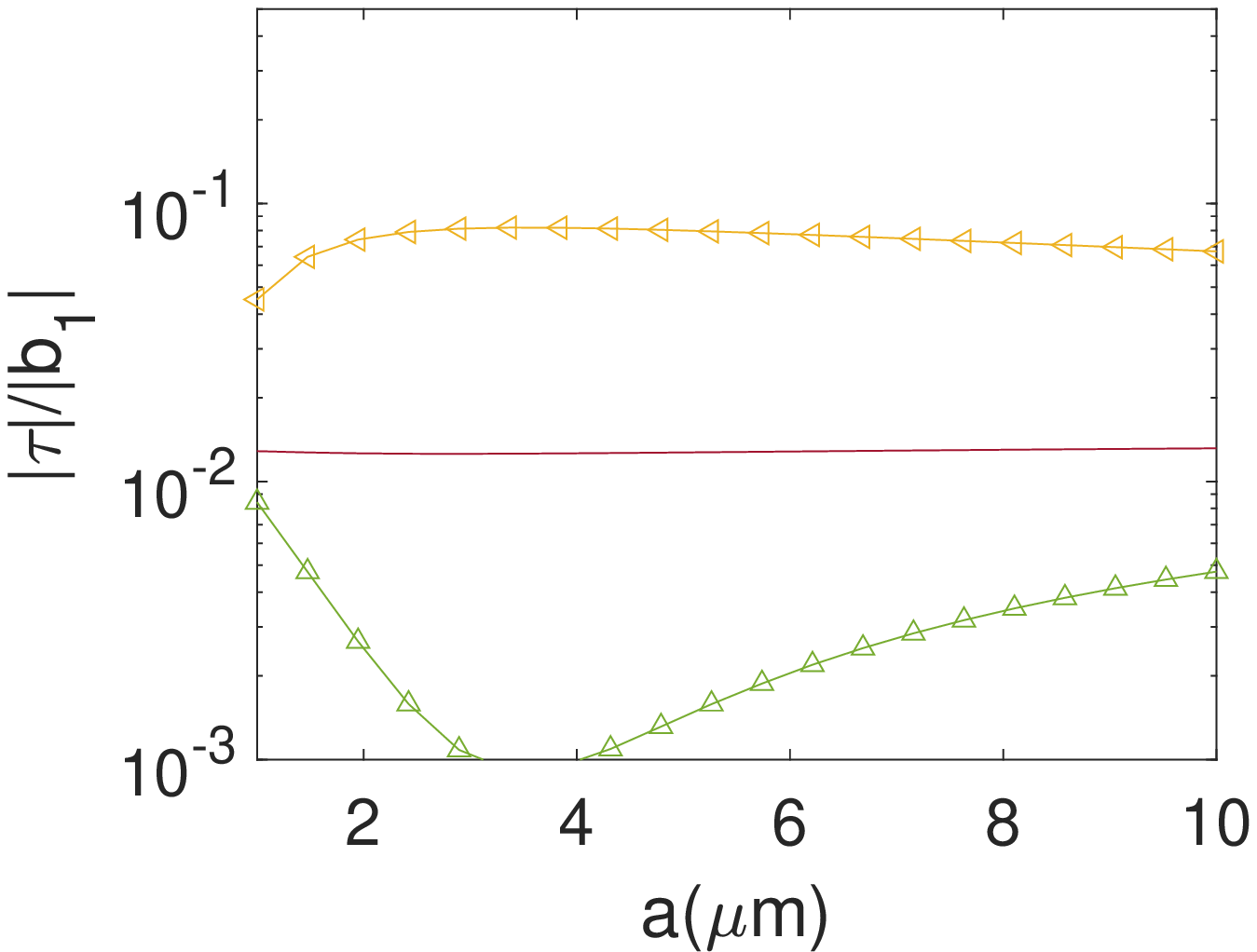}
 \caption{ }
 \label{fig:Fourier52}
\end{subfigure}
 \caption{Fourier decomposition of the Casimir torque for  (a) the homochiral case, and (b) the heterochiral case. The thickness of the crystals is 5 microns.  }\label{fig:Fourier5Micron}
\end{figure}

\begin{figure}[!htbp]
\begin{subfigure}[t]{0.45\textwidth}
\centering
\includegraphics[width=\linewidth]{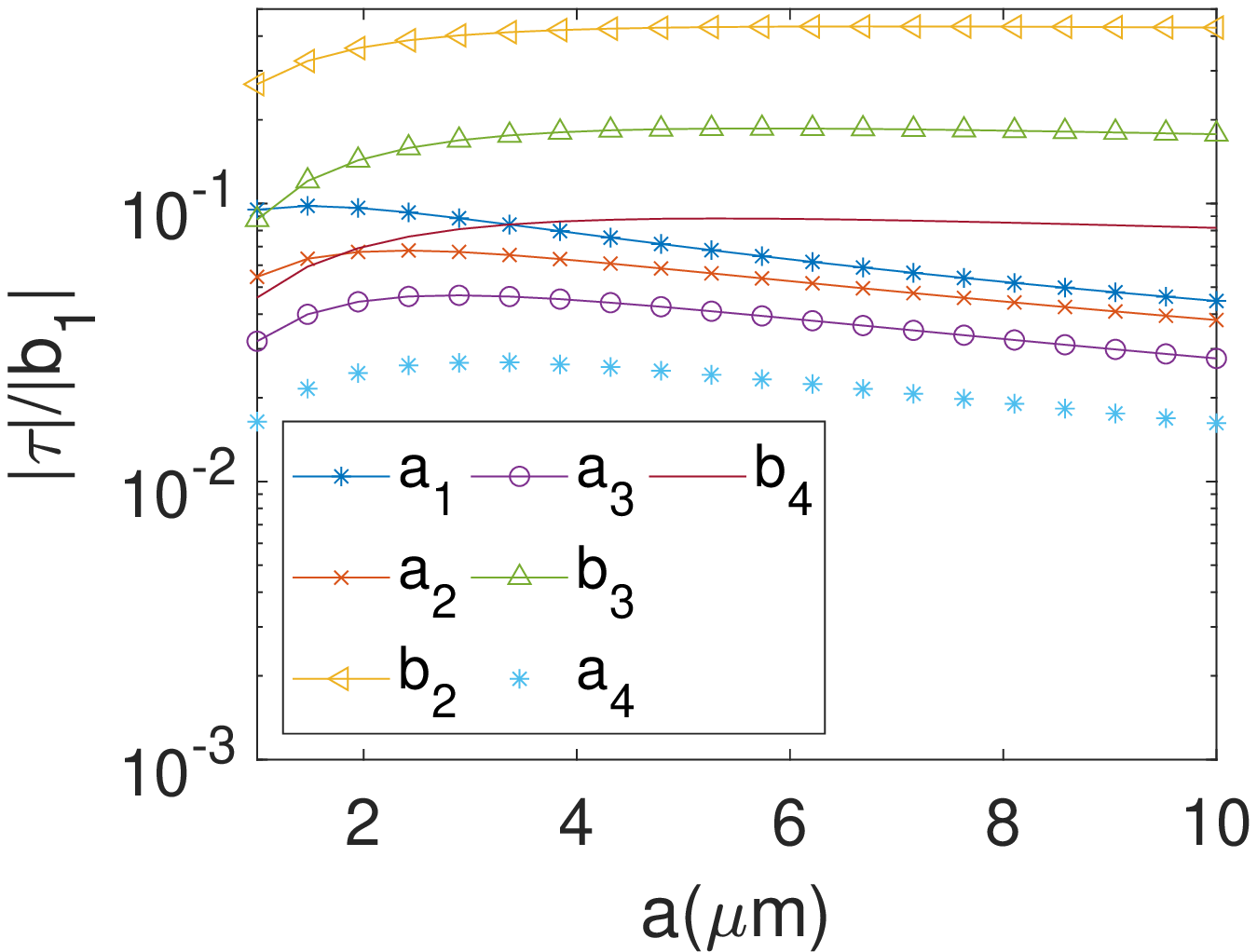}
 \caption{ }
 \label{fig:ten}
\end{subfigure}
\begin{subfigure}[t]{0.45\textwidth}
\centering
\includegraphics[width=\linewidth]{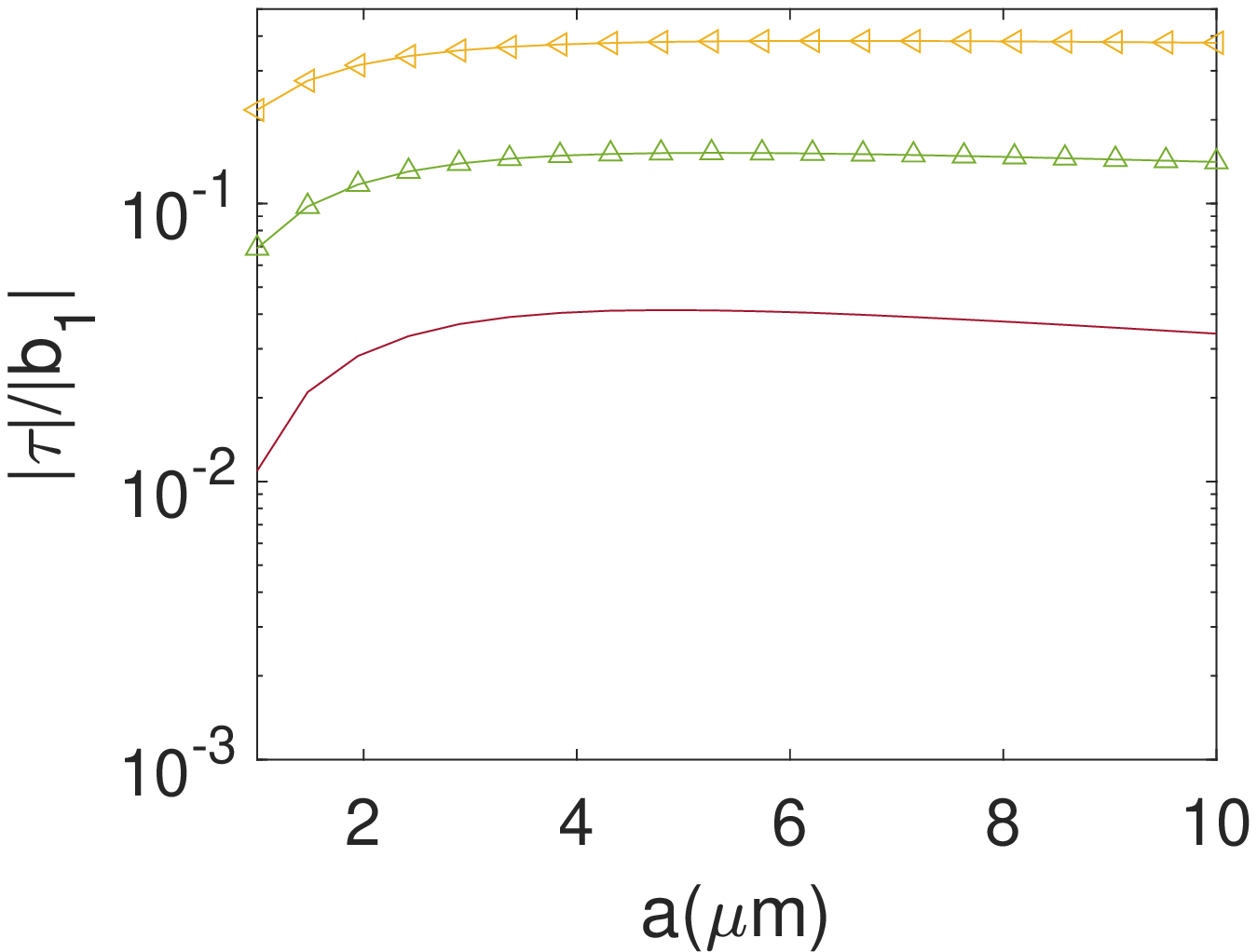}
 \caption{ }
 \label{fig:10}
\end{subfigure}
 \caption{Fourier decomposition of the Casimir torque for  (a) the homochiral case, and (b) the heterochiral case. The thickness of the crystals is 1 micron. }\label{fig:Fourier1Micron}
\end{figure}
Since the torque as a function of the misalignment angle can exhibit deviations from sinusoidal behavior, a natural next step is a Fourier decomposition in terms of this orientational angle.
The Fourier components for a natural number $m$ are given by:

\begin{subequations}
\begin{gather}
     a_m(a)=\frac{2}{\pi}\int\limits_0^\pi \tau(a,\phi)\cos(2m\phi)\ud\phi\\
          b_m(a)=\frac{2}{\pi}\int\limits_0^\pi \tau(a,\phi)\sin(2m\phi)\ud\phi
\end{gather}
\end{subequations}
which will allow us to quantify the deviations from the sinusoidal dependence of the torque on the misalignment angle $\phi$.

The results of the Fourier decomposition for the 5 micron thick crystals can be found in \cref{fig:Fourier5Micron}. As expected, the dominant component is $b_1$, and the other components are plotted as a ratio of this number. It can be seen that for the homochiral case, the component $b_2$ remains constant at the order of 10 \%, while $a_1$ starts at a comparable value which decreases as a function of the separation $a$. The other components are of the order of a few percent or smaller. In the heterochiral case, the higher order components contribute significantly less than in the homochiral case, $b_2$ being the largest component ($<$10\%).

The higher order Fourier components for the 1 micron thick crystals are shown in \cref{fig:Fourier1Micron}. It is clear that the higher order components contribute considerably more than in the case of the thicker crystal. In the homochiral configuration, the component $b_2$ reaches values of up to 43 \% while $b_3$ increases to up to 18\%. The other components also contribute but typically less than 10 \%. In the heterochiral case, $b_2$ and $b_3$ contribute comparably, (38\% and 14\% respectively), but the other components do not play a major role.

\section{Summary and Outlook}

We have investigated the combination of the effects of chirality and finite thickness on the Casimir-Lifshitz torque between cholesteric liquid crystals in the separation range between 1 and 10 microns. We have found that finite thickness effects are associated with deviations from the usual sinusoidal behavior of the Casimir torque as a function of the misalignment angle. 
This can be clearly represented by the higher order orientational Fourier components having a larger relative contribution in the case of thinner crystals. The Fourier decomposition also revealed that the heterochiral case behaves more sinusoidally than the homochiral case.
Surprisingly, the finite thickness increases the torque amplitude at short distance but it decreases the amplitude more quickly as a function of separation. 

Future endeavors could include extending the spiral staircase model to different phases of liquid crystals, for example lyotropic or smectic phases. This would be a challenging problem because it involves nano-confined water \cite{Fumagali2018}. Furthermore, the spiral staircase model may prove useful in other physical contexts where electromagnetic wave propagation through cholesterics plays a role.

\pagebreak
\appendix

\section{Transfer matrix method}\label{TMM}

For the sake of completeness we will briefly describe the transfer matrix method here. For more details about the transfer matrix formalism in general we refer to Refs. \cite{Berreman1972,Yeh1979,Lekner1991}.

We start with a planar multilayer geometry of possibly dielectrically anisotropic materials. The electric permittivity contains two independent elements, $\varepsilon_x$ and $\varepsilon_y$. The uniaxial anisotropy of the material is assumed to lie in the plane of reflection, the laboratory $xy-$plane.  Let the optic axis be rotated by an angle $\theta$ with respect to the laboratory $x-$axis. Then the permittivity is given by

\begin{equation}\label{eq:perm}    \underline{\pmb{\varepsilon}}=\underline{\pmb{R}}\diag(\varepsilon_x,\varepsilon_y,\varepsilon_y)\underline{\pmb{R}}^{-1}
\end{equation}
where $\underline{\pmb{R}}$ denotes the matrix representation of a rotation of $\theta$ around the $z-$axis. The dielectric displacement is $\pmb{D}=\underline{\pmb{\varepsilon}}\cdot\pmb{E}$ and it can be assumed that $\pmb{H}=\pmb{B}$, since the material is not magnetic.

The main idea of the transfer matrix formalism is to treat the Maxwell equations as an eigenvalue problem. Hence we write the vectorial Maxwell equations as four-dimensional eigenvalue equation in layer $j$ of the planar multilayer geometry:

\begin{equation}\label{eq:Maxwell}
\begin{rcases}
\mathbf{k}_j\times\mathbf{E}_j=\frac{\omega}{c}\mathbf{H}_j\\ \mathbf{k}_j\times\mathbf{H}_j=-\frac{\omega}{c}\mathbf{D}_j
\end{rcases}
\rightarrow{\underline{\pmb{Q}}_j \pmb{\Psi}=q ~\pmb{\Psi},}
\end{equation}
where $\pmb{\Psi}=(E_x,E_y,H_x,H_y)^T$, $q$ is a scalar, and $\underline{\pmb{Q}}_j$ is a $4\times4$ matrix with the  respective ordinary and extraordinary eigenvalues

\begin{subequations}
\begin{gather}
\label{eq:qjo}q^\pm_{jo}= \pm \sqrt{\varepsilon_{jy}\zeta^2/c^2+k_\rho^2}\\
\label{eq:qje}q^\pm_{je}= \pm \sqrt{\varepsilon_{jx}\mu_{jy}\zeta^2/c^2+(\varepsilon_{jx}/\varepsilon_{jz})k_\rho^2\cos^2(\theta_j-\eta)+k_\rho^2\sin^2(\theta_j-\eta)}.
\end{gather}
\label{eq:eigenvalues}
\end{subequations}
Since these eigenvalues are distinct, $\underline{\pmb{Q}}_j$ is diagonalizable. So it possible to construct an invertible matrix $\underline{\pmb{S}}_j$ such that 

\begin{equation}
   \underline{\pmb{Q}}_j= \underline{\pmb{S}}_j\cdot\diag(q_{jo},-q_{jo},q_{je},-q_{je}) \underline{\pmb{S}}^{-1}_j,
\end{equation}
where the columns of $\underline{\pmb{S}}_j$ are the eigenvectors of \cref{eq:Maxwell}:

\begin{equation}
                     \underline{\pmb{S}}_j=  \begin{pmatrix}
                              \vertbar & \vertbar & \vertbar & \vertbar\\
                              \pmb{\Psi}_{je}^+ & \pmb{\Psi}_{je}^- & \pmb{\Psi}_{jo}^+ & \pmb{\Psi}_{jo}^-\\
                              \vertbar & \vertbar & \vertbar & \vertbar
                             \end{pmatrix}.
\label{eq:Sj}
\end{equation}

Let the EM wave propagate through a layer of thickness $d_j$. Then a \emph{transfer matrix} $\writematrix{T}_j$ can be defined such that

\begin{equation}
    \writematrix{T}_j\cdot\pmb{\Psi}(z)=\pmb{\Psi}(z+d_j)
\end{equation}
It turns out that $\writematrix{T}_j$ is an exponential of a matrix \cite{Yeh1979,Veble2009}:

\begin{equation}
    \writematrix{T}_j=\exp\left(-i\writematrix{Q}_jd_j\right)
\end{equation}
Because $\writematrix{Q}_j$ is diagonalizable this can be written as

\begin{equation}
    \writematrix{T}_j=\writematrix{S}_j\cdot\diag\left(\exp(-iq_{je}d_j),\exp(iq_{je}d_j),\exp(-iq_{jo}d_j),\exp(iq_{jo}d_j)\right)\writematrix{S}^{-1}_j.
\end{equation}
Now the transfer matrix corresponding to propagation through all $N$ layers is the product of all the individual transfer matrices:

\begin{equation}\label{eq:T_N}
    \writematrix{T}_N=\prod\limits_{j=1}^N\writematrix{T}_j,
\end{equation}
where the product symbol implies matrix multiplication to the right. It is important to make sure the multiplication order is equal to the physical order of the layers, because this affects the end result. \cref{eq:T_N} is a valid representation in the laboratory $xy-$basis. However, we are ultimately interested in the transfer matrix in the $sp-$basis because that is where the Fresnel reflection matrices in \cref{eq:Lifshitz} are given. Hence to obtain the total transfer matrix $\writematrix{M}$ we need to change the basis to the $sp-$basis via

\begin{equation}\label{eq:M}
    \writematrix{M}=\writematrix{S}_0^{-1}\writematrix{T}_N\writematrix{S}_0,
\end{equation}
where the columns of $\writematrix{S}_0$ are the $sp-$eigenvectors corresponding to an isotropic medium:

\begin{equation}
\underline{\pmb{S}}_0=\begin{pmatrix}
                              \vertbar & \vertbar & \vertbar & \vertbar\\
                              \pmb{\Psi}_{js}^+ & \pmb{\Psi}_{js}^- & \pmb{\Psi}_{jp}^+ & \pmb{\Psi}_{jp}^-\\
                              \vertbar & \vertbar & \vertbar & \vertbar
                             \end{pmatrix}.
\label{eq:S0}
\end{equation}

\section{Eigenvectors of Maxwell equations}
The matrix $\underline{\pmb{Q}}_j$ is a 4$\times$4 anti-diagonal block matrix

\[\underline{\pmb{Q}}_j=\begin{psmallmatrix}
    \underline{\pmb{0}}&\underline{\pmb{Q}}_{aj}\\
    \underline{\pmb{Q}}_{bj}&\underline{\pmb{0}}
\end{psmallmatrix},\]

whose non-zero quadrants are given by
\[\underline{\pmb{Q}}_{aj}=\]
\[
\begin{psmallmatrix}
			\tfrac{\omega}{c}\sin(\theta_j-\eta)\cos(\theta_j-\eta)(\mu_{jx}-\mu_{jy})&-k_\rho^2\frac{c}{\varepsilon_{jz}\omega}{+}\tfrac{\omega}{c}(\mu_{jy}\cos^2(\theta_j-\eta)+\mu_{jx}\sin^2(\theta_j-\eta))\\
-\tfrac{\omega}{c}(\mu_{jx}\cos^2(\theta_j-\eta)+\mu_{jy}\sin^2(\theta_j-\eta))&\tfrac{\omega}{c}\sin(\theta_j-\eta)\cos(\theta_j-\eta)(\mu_{jy}-\mu_{jx})
			\end{psmallmatrix}
\]
and

\[
\underline{\pmb{Q}}_{bj}=\]
\[\begin{psmallmatrix}
	\tfrac{\omega}{c}\sin(\theta_j-\eta)\cos(\theta_j-\eta)(\varepsilon_{jy}-\varepsilon_{jx})&-\tfrac{\omega}{c}(\varepsilon_{jy}\cos^2(\theta_j-\eta)+\varepsilon_{jx}\sin^2(\theta_j-\eta)){+}k_\rho^2\tfrac{c}{\mu_{jz}\omega}\\
	\tfrac{\omega}{c}(\varepsilon_{jx}\cos^2(\theta_j-\eta)+\varepsilon_{jy}\sin^2(\theta_j-\eta))&\tfrac{\omega}{c}\sin(\theta_j-\eta)\cos(\theta_j-\eta)(\varepsilon_{jx}-\varepsilon_{jy}).
\end{psmallmatrix},
\]
{where $k_\rho$  and $\eta$ denote the radial and azimuthal components of the wavevector, respectively.}  The EM field mode eigenvectors of $\underline{\pmb{Q}_j}$, obtained from Eq. (4) from the main text, characterized by the  respective subscripts $e$ and $o$, are 

\begin{subequations}
\begin{gather}
\pmb{\Psi}_{jo}^{\pm}=N_{jo}^{-1}
\begin{psmallmatrix}
	\mp\mu_{jy} q_{jo}\frac{\omega}{c}\sin(\theta_j-\eta)\\
	\pm{\mu_{jy}q_{jo}\frac{\omega}{c}\cos(\theta_j-\eta)}\\
	k_{jz}^2\cos(\theta_j-\eta)\\
	-\varepsilon_{jy}\mu_{jy}\frac{\omega^2}{c^2}\sin(\theta_j-\eta)
\end{psmallmatrix},\\
\pmb{\Psi}_{je}^\pm=N_{je}^{-1}
\begin{psmallmatrix}
	-k_{jz}^2\cos(\theta_j-\eta)\\
	-\varepsilon_{jy}\mu_{jy}\frac{\omega^2}{c^2}\sin(\theta_j-\eta)\\
	\pm\varepsilon_{jy}q_{je}\frac{\omega}{c}\sin(\theta_j-\eta)\\
	\mp\varepsilon_{jy}q_{je}\frac{\omega}{c}\cos(\theta_j-\eta)
\end{psmallmatrix},
\end{gather}\label{eq:vec_eo}
\end{subequations}
where $N_{je}$ and $N_{jo}$ denote normalization constants. The matrix $\underline{\pmb{S}}_j$ that changes the from the eigenmode basis to the laboratory $xy-$basis is hence given by
\begin{equation}
\underline{\pmb{S}}_j=
\begin{psmallmatrix}
		-k_{jz}^2\cos(\theta_j-\eta) & -k_{jz}^2\cos(\theta_j-\eta) & -\mu_{jy} q_{jo}\frac{\omega}{c}\sin(\theta_j-\eta) & \mu_{jy} q_{jo}\frac{\omega}{c}\sin(\theta_j-\eta) \\
	-\varepsilon_{jy}\mu_{jy}\frac{\omega^2}{c^2}\sin(\theta_j-\eta) & -\varepsilon_{jy}\mu_{jy}\frac{\omega^2}{c^2}\sin(\theta_j-\eta) & \mu_{jy}q_{jo}\frac{\omega}{c}\cos(\theta_j-\eta) & -{\mu_{jy}q_{jo}\frac{\omega}{c}\cos(\theta_j-\eta)}\\
	\varepsilon_{jy}q_{je}\frac{\omega}{c}\sin(\theta_j-\eta) & -\varepsilon_{jy}q_{je}\frac{\omega}{c}\sin(\theta_j-\eta) & k_{jz}^2\cos(\theta_j-\eta) & k_{jz}^2\cos(\theta_j-\eta)\\
	-\varepsilon_{jy}q_{je}\frac{\omega}{c}\cos(\theta_j-\eta) & \varepsilon_{jy}q_{je}\frac{\omega}{c}\cos(\theta_j-\eta) & -\varepsilon_{jy}\mu_{jy}\frac{\omega^2}{c^2}\sin(\theta_j-\eta) & -\varepsilon_{jy}\mu_{jy}\frac{\omega^2}{c^2}\sin(\theta_j-\eta)
\end{psmallmatrix}.
\label{eq:SDef}
\end{equation}
By the same token, the matrix that changes from the laboratory $xy-$basis to the $sp-$eigenmode basis, is

\begin{equation}
\underline{\pmb{S}}_0=
						\begin{pmatrix}
							0 & 0 & \tfrac{k_3c}{\varepsilon_3\omega} & \tfrac{k_3c}{\varepsilon_3\omega}\\
							1& 1& 0 & 0\\
							\tfrac{k_3c}{\omega} & -\tfrac{k_3c}{\omega} & 0 &0 \\
							0 & 0 & -1 & 1
						\end{pmatrix}.
\label{eq:S0Ex}
\end{equation}
\bibliography{Casimir3}
\end{document}